\documentstyle[epsfig]{aipproc}

\def\lta{\;\rlap{\lower 2.5pt                       
             \hbox{$\sim$}}\raise 1.5pt\hbox{$<$}\;}

\def\msun{M_{\odot}}

\begin{document}
\title{Observations of Type I Bursts from Neutron Stars}

\author{Jean H. Swank}
\address{Laboratory for High Energy Astrophysics\\NASA/GSFC Greenbelt, MD 
20771
}

\maketitle

\begin{abstract}
Observations of Type I X-ray bursts have long been taken as evidence
that the sources are neutron stars. Black body models approximate the
spectral data and imply a suddenly heated neutron star cooling over
characteristic times of seconds to minutes. The phenomena are
convincingly explained in terms of nuclear burning of accreted gas on
neutron stars with low mass companion stars. Prospects are promising
that detailed theory and data from RXTE and future missions will lead
to better determinations of important physical parameters (neutron
star mass and radius, composition of the accreting gas, distance of
the source). Among the variety of bursts observed, there are probably
representatives of different kinds of explosive burning. RXTE's
discovery of a 2.5 ms persistent coherent period from one Type I
burster has now linked bursters indisputably to the epitome of a
neutron star, a fast spinning magnetic compact object. Oscillations in
some bursts had already been thought to arise from the neutron stars'
rotations. Detailed observations of these oscillations are touchstones
of how the explosive bursts originate and progress, as well as
independent measures of the neutron star parameters.

\end{abstract}

\section*{Introduction}

Bursts represent the nuclear burning energy of material accreted on
neutron stars, which is only about 7\% of the gravitational energy
irradiated. But, concentrating many hours worth in 10 seconds, they are
impressive in making the neutron star glow as brightly as the Crab
Nebula for a few seconds. During those seconds it dominates the scene
and we can get a good look. The scene is changing rapidly, so only
observations with large area and high time resolution have a hope of
detecting the kind of dynamical phenomena that could occur. Now that
neutron stars are being studied with these capabilities, we are both
confirming speculations and seeing new aspects we did not envision. I
will briefly summarize the kind of data that have substantiated our
current picture and how the new observations compare. In this, I will
concentrate on the spectral measurements, since the questions
of the burst energy output, the durations, and the correlations with
the persistent flux are well covered in Bildsten's paper in this volume.
Then I will discuss the new results about the 
spins of the 
neutron stars in low-mass X-ray binaries. Fuller details of both 
the earlier and the recent work 
may be found in recent reviews \cite{LvPT95,Bildsten9698,Stroh98,Stroh00}.

\section*{Brief Update on the Spectral Observations}
 
\subsection*{Burst Spectra}

A key indicator that bursters are neutron stars was the nature of the
energy spectra. Of the three limiting types of X-ray spectra,
non-thermal, optically thin high temperature gas, and high temperature
optically thick surface, the spectra of bursts were a good 
match to the last
of these and the parameters made good sense. Projected areas were
within a factor of two of the canonical neutron star and peak
luminosities were within factors of 2 of the Eddington limit for
hydrogen or perhaps helium, for sources concentrated within the
galactic bulge \cite{vP78}. 
At first, only for long bursts and
slow evolution
(100--1000 s), was it possible to say that the black body model was
definitely a better fit than other simple models \cite{Swank77}. 
Now it is possible
for short (10 s) bursts. It still remains true that for spectra of
intervals of only a few tenths of a second, the black body model gives 
a good fit to the data. 

Theoretically, modifications of the spectrum due to scattering and to
temperature distribution in the neutron star atmosphere are expected
to change the spectrum, but while the color temperature is strongly
affected, the shape of the spectrum is not usually very different
from the black body distribution. The spectra obtained by
OSO--8, SAS--3, EXOSAT, TENMA, and Ginga, time resolved to various extents,
depending on the data modes and the detector area, fit black body
distributions and corrected spectral forms equally well. Conceptually,
the fit temperatures are higher than expected because 
the  emissivity is less than unity. 

The predicted differences are excesses at low energies, below about 2
keV, and at high energies, above about 15 keV. London, Taam, and Howard 
\cite{LTH84} first 
included all the most important effects, the radiative transfer, the
competition of scattering and absorption, and inelastic scattering
and coined the phrase "hardening factor". 
Comparisons of burst
spectra with absorbed black body spectra have shown deviations
suggestive of these effects, although usually with detectors where
the response is more uncertain in those ranges, and often when 
fast changes in the burst made it possible that the spectrum changed during 
the interval. 

A potentially very informative deviation from the 
black body form was the dip seen in
TENMA selected burst spectra of 4U~1636--536 \cite{Waki84}, 
4U~1608--522,
and 4U~1746--217, 
at 4.1 keV (in most cases).
If the
energy were redshifted Fe $K_\alpha$, it implies a very large redshift
and puts a strain on the interpretation. That the energy is close to
detector features makes it more difficult to study. It has also been
suggested that the cross-over between a low temperature spectral
component and a higher temperature one, could cause an apparent dip.
Of the current missions, those with modest spectral resolution instruments 
($E/\Delta E > 10$) have low enough area to limit their constraints
and the CCDs have pile-up problems with intense bursts. RXTE has the
requisite area, but is a xenon detector again. It is not yet clear 
whether Chandra, XMM, or Astro-E can address this feature in bursts. 

Typical RXTE PCA spectra for a few tenths of a second, with some exceptions,
are still good
fits to black body spectra, as shown in Figure  1.
Additions of 
spectra to test for theoretical spectral differences have not 
been reported.

\begin{figure}

\centerline{\epsfig{file=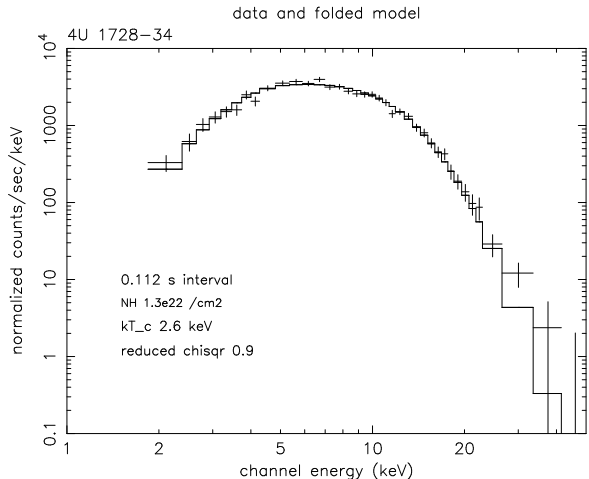,height=2.5in}}
\caption{Spectrum and Fit of a Burst Interval. These 
spectra cannot test the low energy predictions. High energy deviations
require using more data. }
 \label{fig1}

\end{figure}

\subsection*{Color Temperatures and Apparent Radii}

The parameters for flux, temperature, and apparent radius of these
solutions are well determined (except when the color temperature puts
the flux below the low energy threshold of the detectors), as shown
for the example in Figure 2 for a burst with moderate radius
expansion. The temperature starts down after the initial rise, the
radius continues to increase, and the flux has a plateau. These are
consistent with the scenario of the flux rising to the Eddington limit
and lifting surface layers, such that the photosphere still emits the
Eddington limit of radiation while expanding beyond the neutron star.
Some bursts show evidence for both the
Eddington limits of cosmic abundance gas and of
hydrogen-poor/helium-rich gas. Early in the decay the photosphere
contracts and the radius is approximately asymptotically consistent
with that of the neutron star. The asymptotic apparent radius is
not exactly  constant, but slowly climbs as the flux and the
temperature go down, possibly as flux dependent corrections change. 
The radius expands to
around 16 km, less than a factor of two.  
This may be enough, however, for the photosphere
to run into the inner part of the accretion disk, unless the inner part
has been dispersed by the burst already.
For other bursts the
radius can be followed out to 100 km. Even greater expansions have
been seen in rare long superbursts for which the average photon
energy goes below the range of the detector and the rise looks like a
precursor.

\begin{figure}

\centerline{\epsfig{file=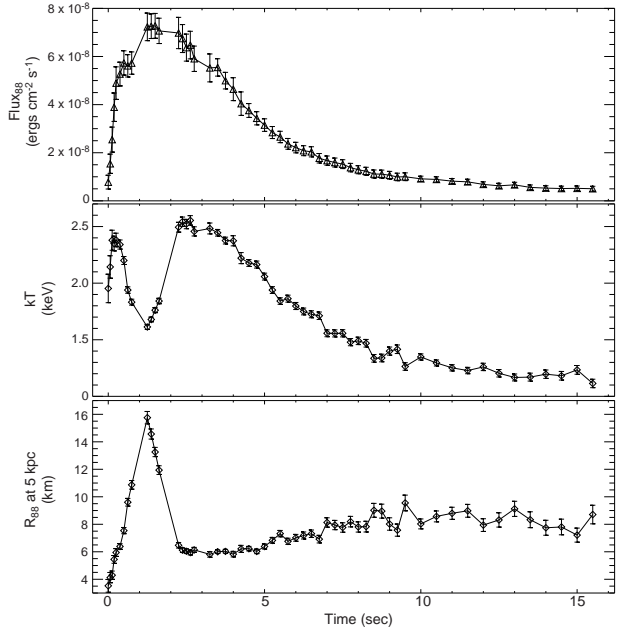,height=3.3in}}
\caption{Evolution of Burst Flux, Temperature and Radius.
The temperature and radius are for simple black body fits.
The data mode gave rise to the gaps.}
\label{f:swank:2}

\end{figure}

\subsection*{Atmospheric Models and Neutron Star Parameters}

It was soon recognized that, in the rather high
temperature atmosphere, scattering would dominate over free-free
absorption and the emissivity would be energy and temperature
dependent. The results of assuming a non-relativistic spherically
symmetric neutron star led to results unlikely to be exactly correct,
radii too small, luminosities exceeding the Eddington limit for He,
and radius variations during bursts decays. The gravitational redshift
effects are straight forward, at least if the star is not rotating
near breakup. Successive calculations of the theoretical
spectra increased in sophistication and applicability. 
\cite{Ebisuzaki87,Titarchuk88}.
They still depend
on the assumptions that the spectrum is formed in an atmosphere in radiative
and hydrostatic equilibrium. 

Ebisuzaki and Nakamura \cite{EN88} used the fact that 
the spectra calculated for a
given luminosity and the spectra observed at a given luminosity fit a
black body distribution for the ``color'' temperature), 
with the
effective temperature a lower value.
The dependence of the  hardening
factor ($T_c/T_{eff}$)
on the ratio of the luminosity to the Eddington luminosity, was 
mapped empirically, 
for various atmospheric constituents and burning scenarios. 
In practice it was
possible, in
the case of radius expansion bursts, to identify the point in the
decay at which for radius expansion bursts, the crust settled back
onto the neutron star. That flux should correspond to the
Eddington limit. Further, the observational information of the fluxes as a
function of time, together with the fit color temperatures, 
allow determination of
the temperature that the emission would have had on the surface at the
touchdown moment. The latter should be a specific function of the mass
and radius and provide one locus of acceptable values. 
For sources
with known distance, the comparison of the observed flux and the
theoretical Eddington limit luminosity give a second locus, to 
uniquely determine the mass and radius.

Efforts to fit EXOSAT data to this picture were plagued by statistical
uncertainties in the determinations \cite{Damen90}. A long burst observed
with Ginga,
exhibited complications in the spectra \cite{vP90}. 
The RXTE PCA
has now obtained
data which is suitable to use for such determinations. However 
a strategy is needed 
to deal with the question of the
fraction of the neutron star that is emitting and the asymmetry of the
emission. The ideas discussed by Bildsten, and the evidence from
the bursts oscillations (discussed below) that the neutron star is not
uniformly emitting (although only sometimes), show that these effects
need to be considered.

\subsection*{Directions of Current Explorations}

There are clearly several regimes of Type I bursts. Differences in
characteristic recurrence times, the ratio $\alpha$ of the 
persistent to burst luminosity, and burst duration, are
observed, not just between sources, but for a given
source. Theoretically, as Bildsten has discussed, these imply
different fuel during the bursts and for accretion of given
abundances close to cosmic,  depend on the mass accretion rate per
unit area. While the main differences are between sources with 5--10 s
bursts, separated by 3--5 hours, versus 100 s bursts separated by 12 or
more hours, there are two less common types of bursts. Some anomalous faint 
bursts have now been confirmed \cite{Gotthelf97,Tomsick99}. 
In addition, long duration
bursts are seen which
last 100--1000 s and have long, but unknown, recurrence times.

The observations are not consistent if the
accretion rate is the only independent parameter. Bildsten proposes that the
reason for correlations opposite those expected for a spherically
symmetric distribution of the burning fuel on a neutron star, may be
that the fuel is not distributed uniformly and the distribution is different
for different states. Taam proposes hysteresis
effects due to changes in internal temperature for changes of
accretion conditions.

For the spectra, persistent flux contributions during the bursts need
to be resolved. Radii of expansion put the photosphere outside where
the kilohertz quasi-periodic oscillations (kHz QPO) 
indicate the inner disk can be (for the atoll
bursters at least). Contributions of reflection, which 
must exist as counterparts
to corresponding optical bursts, need to be taken into account.

\section*{New Information about the Bursters}

\subsection*{A Pulsing Type I Burster}

The Wide Field Camera Experiment on BeppoSAX
has identified at least 16  new bursters, most in the galactic center region.
Most were also transient persistent sources
with peak luminosities less than 100 mCrab.
SAX J1808.5-3658 was one of these \cite{Zand99}. 
The RXTE ASM could see the
transient flux during the 2 week interval that it is was above 20 mCrab.

In May 1998, the RXTE PCA, in slewing between two targets, crossed the
source and flagged that SAX J1808.5-365 was again
bright\cite{Marshall98}. This time the RXTE PCA observed, 
Wijnands \cite{Wijnands98} immediately
found the signal had 401 Hz coherent pulsations and Chakrabarty and
Morgan \cite{CM98} found it had a 2 hour binary Doppler modulation
with a tiny  mass function ($3.8 \times 10^{-5} \msun$).  Consideration
of possible orbits and companions suggested the companion could be
several times less than $\approx 0.15 \msun$ and the orbit possibly viewed nearly face-on.

Although
the flux after 2 weeks dropped within 2 days by a factor of 5 the
energy spectrum did not change and pulsations continued to be
observed. Naively at least, accretion onto the neutron star was able
to occur through the observed range of accretion rate.  Quantitative
limits with various assumptions about the laws describing the gas in
the accretion disk (e.g. cold gas pressure or radiation pressure
supported)  led to estimates of the magnetic field within a
factor of two of $ 5 \times 10^8$ Gauss\cite{Psaltis99}.
The X-ray pulsar's period of 2.5 ms is in the middle of the periods of 
a half dozen
low-mass binary radio pulsars (1.6--7.5 ms) and a magnetic field
$10^8-10^9$ is proximate to the magnetic fields of the radio
pulsars. It appears likely that such a pulsar source
could evolve to a similar radio pulsar.

\subsection*{Flux Oscillations during Bursts}

The dynamical power spectrum of a burst from 4U 
~1728--34 is shown in Figure 3. Strohmayer 
discovered both the kHz QPO 
and the burst 
oscillations in the first observations of this source.
\begin{figure}

\centerline{\epsfig{file=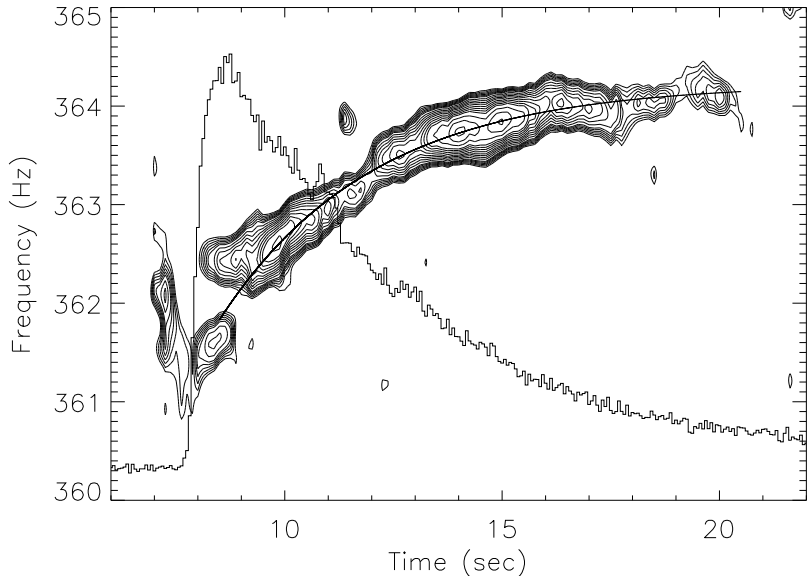,height=3.0 in}}
\caption{Contours of Power Spectral Density versus Time for a Burst
from 4U 1728-34. 
Power is calculated for 2 s intervals starting every 0.25 s.
for smoothed contours. 
An exponential recovery fit to the frequency is superposed 
as well as the light curve.}

\label{fig3}

\end{figure}
The number of sources for which oscillations have been seen during
some bursts is still six, as it was by the end of the first year of
RXTE operations \cite{Stroh98}. 
KHz QPO in the persistent
flux have been seen in a dozen atoll bursters at frequencies of
300-1200 Hz.  There are twice as many bursters with weak persistent
flux. A small number of these have exhibited the  QPO, 
but for many of these
the flux is very weak and the limits are not very constraining. Bursts
have also been seen from them. In some cases it is known that no
oscillations were seen in the bursts. But in others the results are
still pending. No oscillations have been seen in bursts that have been
observed from the Z-sources, GX~17+2 and Cyg~X--2.

Which bursts exhibit oscillations?  Some interesting correlations 
have been noted, but the
question has not yet been addressed systematically. Definitely
some bursts with strong radius expansion do not exhibit oscillations.
On the other hand for bursts with mild radius expansion, there may be
oscillations before the expansion starts and after it has
subsided. This is in agreement with the oscillations being associated
with the neutron star surface.

For the atoll bursters the position in the color-color diagram for the
persistent flux has long
been argued to be response to the accretion rate \cite{vdK90}, exactly because
burst properties correlated with the position. 
One of the properties that could be correlated
is the inhomogeneity that shows up in oscillations.

\subsection*{KiloHertz Quasi-periodic Oscillations in Persistent Flux}

The characteristic pattern of kiloHertz oscillations is two
frequencies 250-350 Hz apart, where the difference stays approximately
constant, while the frequencies can change by a factor of two.  Figure 4
compares the distribution of the differences with the frequencies of
oscillations seen in the bursts. Of the four sources for which the
burst frequency is above 500 Hz, two have the pair of kHz QPO and the
burst frequency is approximately twice the difference frequency. For
the two with the lowest burst frequencies, these are very close to
their difference frequencies.

\begin{figure}

\centerline{\epsfig{file=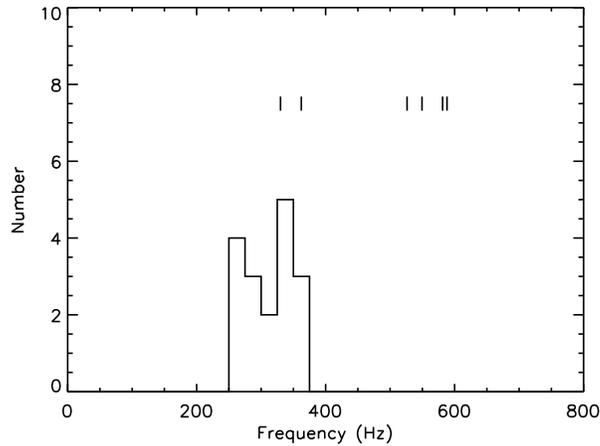,height=2.3in}}
\caption{Histogram of KHz QPO Differences and Burst Oscillation Frequencies.
The frequencies of greatest power during bursts are shown for 
the six bursters, 4U 1728--34, 4U 1636--53, 4U 1702--429, KS 1731--26, 
Aql X-1, and MXB 1743--29, with a histogram of the 
difference between the two kHz QPO in persistent fluxes. 
.}
\label{fig4}

\end{figure}

Models with one of these frequencies being a beat with a lower
frequency would provide a natural explanation. If the burst
oscillations are the neutron star's spin, it provides an obvious
candidate to beat with one of the QPO. In this case, complications
have to explain discrepancies of about 5\%. Miller shows that
the  physical mechanisms in 
the ``sonic-point model'' can accommodate them \cite{Miller00}.

If the difference has nothing to do with the spin, but instead
reflects frequencies entirely in the inner accretion disk, some effect
of these during the burst has to explain the burst
oscillations. Otherwise the closeness of the numbers seems an unlikely
coincidence. The spins could be deduced from more subtle effects on
the Kepler and apsidal motion to make the data fit \cite{Stella99}.

If the differences are the spins, the spins have a narrow distribution
(257-358~Hz, corresponding to 2.8-3.8 ms). The sources have widely
differing accretion rates (currently). Bildsten has explained this
pile-up of the spins as due to the equilibrium between spin-up torque
and gravitational radiation of the neutron star angular momentum
\cite{Bildsten98}. This avoids a conspiracy between the luminosity and 
the magnetic field that is implied for equilibrium accretion at a 
magnetosphere.

If the differences do represent the spins, then the bursts in some
cases show the fundamental and in others the signal is dominated by
the first harmonic. For at least 4U 1636-53, a weak amplitude at the
fundamental appears at the beginning of some bursts, as would be the case
 if the burning
starts at one magnetic pole and very quickly spreads to the optical
pole. 

\subsection*{Neutron Star Spin Periods}

Strohmayer has led investigation of several predictions of the model
in which a hot spot on the neutron star surface causes the flux
oscillations as the star rotates\cite{Stroh00,Strohb00}. 
They have appeared to confirm the
interpretation in terms of the rotation period of the neutron
star. 

(1) In the bursts in which oscillations appear near the beginning of
bursts and then die out, the amplitude of the oscillations can reach
as high as 70 \%. The amplitude dies down as the burst progresses in a
way consistent with the spread of a hot spot. This increase in the
burning area agrees with the spectral analysis.

(2) The asymptotic oscillations during the declining phase of the
burst are usually well defined, suggesting a frequency characteristic
of the neutron star (although the cause of asymmetry in this phase is
not known). For some sources the asymptotic values vary little from
burst to burst, even from year to year, independent of differences in
the bursts. This seemed to have the promise of being a way to discover
the binary motion, because differences corresponded to Doppler motions
of 20 km s$^{-1}$ which would be reasonable for binary motion. However,
as the number of bursts studied has increased, evidence has appeared
for larger differences. 

(3) The oscillations in the burst tails are consistent with being
coherent. There is a characteristic exponential approach to the
asymptote $\nu_a$, with the frequency being a function of time:
$\nu (t) = \nu_a (1-\delta e^{-t/\tau})$ \cite{Stroh99}.
With the frequency variation taken into account, 
the coherence is consistent with the duration of the signal. 

The change in frequency as a function of the decay cries out for a
theory, if the spin of the neutron star is being deduced from it. It
has been pointed out that if the heated crust is lifted enough by the
radiation to be released from viscous drag, so that it can rotate
freely and conserve its own angular momentum, only 10-20 M of lift is
required to explain the 0.3-1 \% change in frequency as the
burst cools and the crust presumably subsides. Bildsten calculated 
such a lift in 
the development of bursts \cite{Bildsten95}.
From this point of view it appears physically
reasonable. However, the implied frequency difference between the neutron
star's interior and the crust, of 1-3 Hz, implies that in a few
seconds the hot spot at the photosphere has rotated several times
around a spot fixed on the neutron star, such as a magnetic pole. If
the magnetic field and the crust material are coupled, the magnetic field
would have been wound up several times, and be an important factor.

\section*{Conclusions}

In Type I bursts, we can truly say that the explosion throws light on
the objects and on the process causing the explosion. Limits on the
neutron star masses, radii, and distances come from the spectra, and
if oscillations are from a rotating hot spot, from the oscillation
amplitudes. At least one burster definitely is a low magnetic field
($10^8 - 10^9$ Gauss) fast rotating pulsar and the evidence is strong
for millisecond periods in 6 other bursters. The spectral and timing
data tell a story of the dynamics of the crust as it is lifted,
expelled sometimes, and settles back down on the neutron star.

If oscillations are found in bursts from a pulsing burster like SAX
J1808.4-3658, comparison to the pulse period would be
very revealing. Finding kHz QPO in the persistent flux from such 
a source would also be revealing, allowing comparison of the difference
frequency with the known pulse frequency. In both cases, it might 
be that the
phenomenon does not occur under the conditions of field and accretion 
rate pertaining for the pulsar. 
It is clear that the phenomena depend on the accretion rate and on the
magnetic field. More work is needed on whether more sources have burst
and persistent flux oscillations and possibly coherent 
pulsations at some level.

The newly found phenomena have the potential for clarifying the
geometry and flow patterns of the accretion and independently
constraining the neutron star parameters. Definitive conclusions about
the burster neutron stars based on the spectra and luminosities have
been hampered by statistical as well as possible systematic errors and
by evidence that over determined results were not consistent. Now that
new degrees of freedom are known (fast spin, asymmetrical temperature
distribution, disk very close to the neutron star), a verifiable
corrected picture should be pursued.

\medskip

\noindent I greatly appreciated the in-depth discussions at the 1999
Aspen Summer workshop ``X-Ray Probes of Relativistic Effects near
Neutron Stars and Black Holes."

\end{document}